\documentclass[conference,a4paper]{IEEEtran}

\usepackage[utf8]{inputenc} 
\usepackage[T1]{fontenc}
\usepackage{url}              % provides \url{...}
\usepackage{cite}             % improves presentation of citations

\usepackage[cmex10]{amsmath}  % Use the [cmex10] option to ensure complicance
\usepackage{mleftright}       % fix to wrong spacing of \left-,
\mleftright                   % \middle- \right-commands 

\usepackage{graphicx}         % provides \includegraphics{...} to
\usepackage{booktabs}         % fixes poor spacing in tables and
\usepackage{PerlazaISIT2023}

\IEEEoverridecommandlockouts
\begin{document}

\title{$\boldmath{2 \times 2}$ Zero-Sum Games with Commitments and  Noisy Observations}

%%%%%%
%\author{%
%  \IEEEauthorblockN{Anonymous Authors}
%  \IEEEauthorblockA{%
%    Please do NOT provide authors' names and affiliations\\
%    in the paper submitted for review, but keep this placeholder.\\
%    ISIT23 follows a \textbf{double-blind reviewing policy}.}
%}

%%%%%% Please only add the author names and affiliations for the FINAL
%%%%%% version of the paper, but NOT for the paper submitted for review!
%
%%%%%
%%%%% Single author, or several authors with same affiliation:
% \author{%
%   \IEEEauthorblockN{Stefan M.~Moser}
%   \IEEEauthorblockA{ETH Zürich\\
%                     8092 Zürich, Switzerland\\
%                     moser@isi.ee.ethz.ch}
%                   }
%
%%%%%
%%%%% Several authors with up to three affiliations:
% \author{%
%   \IEEEauthorblockN{Stefan M.~Moser}
%   \IEEEauthorblockA{ETH Zürich\\
%                     ISI (D-ITET), ETH Zentrum\\
%                     8092 Zürich, Switzerland\\
%                     moser@isi.ee.ethz.ch}
%   \and
%   \IEEEauthorblockN{Albus Dumbledore and Harry Potter}
%   \IEEEauthorblockA{Hogwarts School of Witchcraft and Wizardry\\
%                     Hogwarts Castle\\ 
%                     1714 Hogsmeade, Scotland\\
%                     \{dumbledore, potter\}@hogwarts.edu}
% }

%%%%%   
 \author{%
   \IEEEauthorblockN{Ke Sun\IEEEauthorrefmark{1}, 
   		Samir M. Perlaza\IEEEauthorrefmark{2}\IEEEauthorrefmark{3}\IEEEauthorrefmark{4},
               	Alain Jean-Marie\IEEEauthorrefmark{2}
                 }
\IEEEauthorblockA{\IEEEauthorrefmark{1}%
		    School of Computer Engineering and Science, Shanghai University, 
                     Shanghai, China}
   \IEEEauthorblockA{\IEEEauthorrefmark{2}%
                     INRIA, Centre Inria d'Université Côte d'Azur,
                     Sophia Antipolis, France.}
   \IEEEauthorblockA{\IEEEauthorrefmark{3}%
                     ECE Dept. Princeton University, 
                     Princeton, 08544 NJ, USA.}
   \IEEEauthorblockA{\IEEEauthorrefmark{4}%
                     GAATI, Université de la Polynésie Française,
                     Faaa, French Polynesia.}
                     \thanks{ This work is supported by the Inria Exploratory Action -- Information and Decision Making (AEx IDEM).  Ke Sun was with INRIA while developing the majority of this work. }
 }

%%%%%% Many authors with many affiliations:
% \author{%
%   \IEEEauthorblockN{Ke Sun\IEEEauthorrefmark{1},
%                    Samir M. Perlaza\IEEEauthorrefmark{2},
%                    and Alain Jean-Marie\IEEEauthorrefmark{2}}
%   \IEEEauthorblockA{\IEEEauthorrefmark{1}%
%                     School of Computer Engineering and Science, Shanghai University, 
%                     Shanghai, China}
%   \IEEEauthorblockA{\IEEEauthorrefmark{2}%
%INRIA, Centre Inria d'Université Côte d'Azur, Sophia Antipolis, France
%                 }
%
%    \thanks{This work was supported by the Inria Exploratory Action – Information and Decision Making (IDEM).}
% }

\maketitle

%%%%%
%% Abstract: 
%% If your paper is eligible for the student paper award, please add
%% the comment "THIS PAPER IS ELIGIBLE FOR THE STUDENT PAPER
%% AWARD." as a first line in the abstract. 
%% For the final version of the accepted paper, please do not forget
%% to remove this comment!
%%

\begin{abstract}\boldmath{
In this paper,~$2\times2$ zero-sum games are studied under the following assumptions: 
$(1)$~One of the players (the leader) commits to choose its actions by sampling a given probability measure (strategy); 
$(2)$~The leader announces its action, which is observed by its opponent (the follower) through a binary channel; and 
$(3)$~the follower chooses its strategy based on the knowledge of the leader's strategy and the noisy observation of the leader's action. 
Under these conditions, the equilibrium is shown to always exist. Interestingly, even subject to noise, observing the actions of the leader is shown to be either beneficial or immaterial for the follower. More specifically, the payoff at the equilibrium of this game is upper bounded by the payoff at the Stackelberg equilibrium (SE) in pure strategies; and lower bounded by the payoff at the Nash equilibrium, which is equivalent to the SE in mixed strategies.
Finally, necessary and sufficient conditions for observing the payoff at equilibrium to be equal to its lower bound are presented. Sufficient conditions for the payoff at equilibrium to be equal to its upper bound are also presented.
}
\end{abstract}
 \vspace{-0.24em}
\section{Introduction}
 \vspace{-0.1em}
Zero-sum games (ZSGs) are mathematical models describing the interaction of mutually adversarial decision makers. 
%%
%In the realm of machine learning, ZSGs  have played a central role in the development of techniques such as generative adversarial networks (GANs) \cite{goodfellow_2014_generative}; adversarial training \cite{madry_2017_towards}; and evaluation of generalization capabilities of learning algorithms \cite{mustafa_2022_generalization}.
%Within this context, 
Two solution concepts are often adopted for predicting the outcome of ZSGs: the Nash equilibrium (NE)~\cite{nash_1950_equilibrium} and the Stackelberg equilibrium (SE) \cite{Stackelberg-1952}. %Such predictions occur under different assumptions. 
The NE is a prediction observed under the assumption that both players simultaneously choose their strategies (probability measures over the set of possible actions). %Hence, at an  NE, decision makers make their decisions (take their actions) by sampling  probability measures that lead to expected payoffs that are optimal given the opponents' strategies. %That is, at an NE, none of the decision makers benefits from using an alternative strategy.
On the other hand, the SE describes the outcome in which one of the players (the leader) commits to use a particular strategy before its opponent (the follower). In such a case, the follower chooses its strategy as a best response to the commitment of the leader. 
Commitments are said to be in mixed strategies when the leader is allowed to commit to strategies whose support contains more than one action. In this case, the relevant solution concept is the SE in mixed strategies~\cite{conitzer_2006_computing,  conitzer_2016_stackelberg,  leonardos_2018_commitment, von_2010_leadership}. 
Interestingly, in ZSGs, the payoffs at the NE and the SE in mixed strategies are identical, as shown in~\cite{v1928theorie}. %Thus, commitments in mixed strategies in ZSGs do not represent particular benefits for either player.
The commitment is said to be in pure strategies when the leader is constrained to commit to play one action with probability one. This is assimilated to the case in which the follower perfectly observes the action played by the leader.
The relevant solution concept under these assumptions is the SE in pure strategies~\cite{Stackelberg-1952, simaan_1973_stackelberg, simaan_1973_additional}.
The expected payoff at the SE in pure strategies is equal to the~$\min\max$ or~$\max\min$ solution, where the optimization is over the set of actions \cite{jin_2020_local, bai_2021_sample}. In this case, the payoff at the SE in pure strategies might be significantly different from the payoff at the NE.
%
%In adversarial training, the underlying assumption is that the follower (the attacker or adversary) perfectly observes the action played by the leader (the learner) \cite{huang_2022_robust, zuo_2021_adversarial, bruckner_2011_stackelberg, gao_2022_achieving, bai_2021_sample}. Similarly, in data integrity attacks, the follower (the learner) perfectly observes the action of the leader (the attacker) \cite{chivukula_2017_adversarial, liu_2009_game, kantarciouglu_2011_classifier}. 
%That is, adversarial training and data integrity attacks are studied using the SE in pure strategies.
%%
%Alternatively, GANs are modelled by ZSGs in which the relevant solution concept is the NE (or SE in mixed strategies)~\cite{hsieh_2019_finding, oliehoek_2018_beyond}. Essentially, ZSGs are used to predict game outcomes in terms of mixed strategies (probability measures), instead of actions (pure strategies). 

In a nutshell, the underlying assumption of the SE in mixed strategies is that the strategy to which the leader commits to is perfectly observed by the follower and the actions are unobservable.  Alternatively, the assumption of the SE in pure strategies is that actions are perfectly observable, which makes the notion of commitment irrelevant. This is essentially because the follower can always respond with an optimal action to the action played by the leader, regardless of the commitment. 
Nonetheless, often, the actions of the leader are neither unobservable nor perfectly observed. Instead, observations might be obtained subject to noise.
%%
%In real system implementations, the observations of actions are not exempt from such impairments. Nonetheless, 
%The impact of noisy observations of the actions of the leader in ZSGs with commitments remains an uncharted territory. This is in part due to the lack of simple and adapted game solution concepts. This work makes progress in this direction and proposes a ZSG formulation with commitments in which actions are observed subject to noise.
\vspace{-0.2em}
\subsection{Previous Works}
\vspace{-0.2em}
The analysis of noisy observations of the actions played by a leader in ZSGs started in the realm of information theory~\cite{wallmeier_1988_games}. Therein, an external entity referred to as \emph{the informant} observes the action of the leader, encodes it and transmits it through a discrete memoryless channel (DMC) to the follower. The latter decodes the action of its opponent and thus, chooses its own action. In \cite{wallmeier_1988_games}, commitments are not considered and the observation is noisy due to the impairments typical to data-transmission. 
In the realm of game theory, bi-matrix games with commitments and observability started with the work of Bagwell \cite{bagwell_1995_commitment}. Therein, the leader is restricted to commit to a pure strategy, while the follower might observe  a different pure strategy with positive probability. Note that this game is identical to a game without commitments in which the leader plays an action while the follower observes a different action with positive probability before choosing its own action, as described in numerous scenarios \cite{muller_2001_quality, ferreira_2011_note, teng_2013_generalized, van_1997_games, guth_2006_noisy, bizzotto_2022_limits, adolph_1996_commitment}.   
\vspace{-0.2em}
\subsection{Contributions}
\vspace{-0.2em}
For pedagogical purposes, the analysis is restricted to two-player two-action ZSGs, which  capture all interesting challenges due to the noisy observations in the presence of commitments.   
One of the main contributions is a new game formulation in which the follower obtains a noisy observation of the action played by the leader, whereas the commitment is assumed to be perfectly observed.
The game is proved to always possess an equilibrium. The optimal commitments are characterized and the set of best responses of the follower is thoroughly described. An explicit expression for the payoff at the equilibrium is derived. The payoff at equilibrium is greater than the payoff at the NE exclusively when the ZSG exhibits a unique NE in mixed strategies. In all other cases, e.g., ZSG exhibiting strategic dominance, unique NE in pure strategies, or infinitely many NEs, the payoffs with and without observations are identical. When the observation of the action of the leader is noiseless, the payoff at the equilibrium is the same as the payoff at the SE in pure strategies. 

\section{Game Formulation}\label{SecGameFormula}%

Consider a two-player zero-sum game in normal form with a payoff matrix 
\begin{IEEEeqnarray}{rcl}
\label{EqMatrixU}
\matx{u} & = & 
\begin{pmatrix}
u_{1,1} & u_{1,2}\\
u_{2,1} & u_{2,2}
\end{pmatrix}.
\end{IEEEeqnarray}
%Let the elements of the set~$\set{K} \triangleq \{1,2\}$ represent the indices of the players. 
Let the elements of the set~$\set{K} \triangleq \{1,2\}$ represent the indices of the players; and let the elements of the set~$\set{A}_1 = \set{A}_2 \triangleq \{a_1, a_2\}$ represent the actions of the players.
Hence, for all~$(i,j) \inCountTwo^2$, when \Pone plays~$a_i$ and \Ptwo plays~$a_j$, the outcome of the game is~$u_{i, j}$.
\Pone and \Ptwo choose their actions to maximize and minimize their payoffs, respectively.
When players simultaneously choose their actions in the absence of commitments, the game is represented by the tuple 
\begin{IEEEeqnarray}{rcl}
\label{EqTheGame}
\gameNF{\matx{u}} & \triangleq & \left(\set{K}, \set{A}_1 , \set{A}_2 , \matx{u} \right),
\end{IEEEeqnarray} 
and the solution concept is the NE.

When the game is played with commitments and noisy observations, it unfolds in three stages. In the first stage, \Ptwo  announces its strategy to \Pone and commits to choose its actions by using such a strategy.
A strategy for \Ptwo  is a probability measure denoted by~$P_{A_2} \in \simplex{\set{A}_2}$.  
In stage two, \Ptwo plays action~$b \in \set{A}_2$ with probability~$P_{A_2}\left( b \right)$, while 
\Pone observes action~$\tilde{b} \in \set{A}_2$ with probability~$P_{\tilde{A}_2 | A_2 = b} \left( \tilde{b} \right)$. That is, \Pone obtains a noisy observation of the action played by \Ptwo. 
The tuple of probability measures 
\begin{IEEEeqnarray}{rcl}
\label{EqTheRabbitChannel}
P_{\tilde{A}_2 | A_2}  & \triangleq & \left( P_{\tilde{A}_2 | A_2 = a_{1}}, P_{\tilde{A}_2 | A_2 = a_{2}} \right) \in \simplex{\set{A}_2}^{2},
\end{IEEEeqnarray}
 which is a parameter of the game, defines a discrete memoryless channel (DMC) as in \cite{Shannon-1948a, Shannon-1948b}. 
In the final stage, \Pone plays the action~$a \in \set{A}_1$, with probability~$P_{A_1| \tilde{A}_2 = \tilde{b}}\left( a \right)$ and both players obtain their payoffs.

A strategy for \Pone is a tuple of probability measures 
\begin{IEEEeqnarray}{rcl}
\label{EqStratP1}
P_{A_1| \tilde{A}_2} &\triangleq & \left( P_{A_1| \tilde{A}_2 = a_{1}},  P_{A_1| \tilde{A}_2 = a_{2}} \right)  \in  \simplex{\set{A}_1}^{2},
\end{IEEEeqnarray} 
which is chosen based on the commitment (the probability measure~$P_{A_2}$). 
\Pone chooses its action by sampling the probability measure~$P_{A_1| \tilde{A}_2 = \tilde{b}}$, which is conditioned on the noisy observation~$\tilde{b}$.

%Denote by~$A_1$,~$A_2$, and~$\tilde{A}_2$ the random variables representing the actions of \Pone, \Ptwo, and the noisy observation of the action played by \Ptwo, respectively. 

%Let~$P_{A_1 \tilde{A}_2 A_2} \in \simplex{\set{A}_1 \times \set{A}_2 \times \set{A}_2}$ be  the probability measure that satisfies for all~$\left( a,  \tilde{b},b  \right) \in \set{A}_1 \times \set{A}_2 \times \set{A}_2$, 
%
%\begin{IEEEeqnarray}{rcl}
%\label{EqTheJointThing}
%P_{A_1 \tilde{A}_2 A_2} \left( a, \tilde{b},  b \right) = P_{A_2} \left( b \right) P_{\tilde{A}_2 | A_2 = b} \left(  \tilde{b}  \right) P_{A_1 | \tilde{A}_2 = \tilde{b}} \left( a  \right). 
%\end{IEEEeqnarray}
%
The expected payoff obtained by the players is determined by the function~$v:  \simplex{\set{A}_1} ^{2} \times \simplex{\set{A}_2} \to \reals$, such that given the strategy~$P_{A_1| \tilde{A}_2}$ in~\eqref{EqStratP1} of \Pone and the strategy~$P_{A_2}$ of \Ptwo, the expected payoff  is  
\begin{IEEEeqnarray}{rcl}
\nonumber
& & v\left(P_{A_1| \tilde{A}_2}, P_{A_2} \right) \\
\label{EqTheCostFunction}
& = & \hspace{-1.5ex} \sum_{(i,j)\inCountTwo^2 } \hspace{-1.5ex} u_{i,j} \left(\sum_{\tilde{b} \in \set{A}_2} P_{A_1 | \tilde{A}_2 = \tilde{b}} \left( a_i  \right) P_{\tilde{A}_2 | A_2 = a_j} \left(  \tilde{b}  \right)  \right)  P_{A_2} \left( a_j \right). \IEEEeqnarraynumspace \supersqueezeequ
%P_{A_1 A_2} \left( a_{i} , a_{j}\right),  
\end{IEEEeqnarray}
Often, it is said that \Ptwo acts as the leader and \Pone acts as the follower to highlight the order in which players choose their actions. %\footnote{The choice of indices~$1$ and~$2$ for the follower and the leader,  which might appear counter-intuitive, is adopted to ease identifying the connections with the games with incomplete information introduced by \cite{aumann_1995_repeated}.}.

The extension of the game~$\gameNF{\matx{u}}$ in~\eqref{EqTheGame} to capture commitments and noisy observations through the DMC in~\eqref{EqTheRabbitChannel} is represented by the tuple:
\begin{IEEEeqnarray}{rcl}
\label{EqTheNewGame}
\gameNF{\matx{u}, P_{\tilde{A}_2 | A_2}} & \triangleq & \left(\set{K}, \set{A}_1 , \set{A}_2 , \matx{u}, P_{\tilde{A}_2 | A_2} \right).
\end{IEEEeqnarray}

\subsection{Equilibrium}

%\subsubsection{The Set of Best Responses of \boldmath{\Pone}}
The set of best responses of \Pone to the commitment announced by \Ptwo is determined by the correspondence~$\BR_1: \simplex{\set{A}_{2}} \to \mathscr{F}\left(\simplex{\set{A}_1}^{2}\right)$, where $\mathscr{F}\left(\simplex{\set{A}_1}^{2}\right)$ denotes the power set of~$\simplex{\set{A}_1} \times \simplex{\set{A}_1}$. In particular, the set of best responses to the commitment~$P_{A_2}$ is
\begin{IEEEeqnarray}{rcl}\label{Equ:BR_1}
\label{EqBR1}
\BR_1\left( P_{A_2}  \right) & = &\arg\max_{ Q_{A_1|\tilde{A}_2} \in \simplex{\set{A}_1}^{2}} v(Q_{A_1|\tilde{A}_2}, P_{A_2}),  \IEEEeqnarraynumspace 
\end{IEEEeqnarray}
where the function~$v$ is defined in~\eqref{EqTheCostFunction}.
Let the real-valued function~$\hat{v}:\simplex{\set{A}_2} \rightarrow \reals$ be such that 
\begin{IEEEeqnarray}{rcl}
\label{Eqvhat}
\hat{v}\left(P_{A_2} \right) = \max_{Q_{A_1|\tilde{A}_2} \in  \BR_{1}\left( P_{A_2} \right)} v\left( Q_{A_1|\tilde{A}_2}, P_{A_2} \right),
\end{IEEEeqnarray}
where the function~$v$ is defined in~\eqref{EqTheCostFunction}, and the correspondence~$\BR_{1}$ is defined in~\eqref{EqBR1}.
\Ptwo chooses its strategy (commitment)~$P_{A_2}$ assuming that \Pone uses a best response to such strategy. 
Hence, the optimal commitments are the minimizers of~$\hat{v}$ in~\eqref{Eqvhat}.

Equipped with these objects, the solution concept for the game~$\gameNF{\matx{u}, P_{\tilde{A}_2 | A_2}}$ in~\eqref{EqTheNewGame} is the following.

\begin{definition}[Equilibrium]\label{DefEquilibriumTilde}
The tuple~$\left( P_{A_1| \tilde{A}_2}, P_{A_2}\right) \in  \simplex{\set{A}_1}^{2} \times \simplex{\set{A}_2}$ is said to form an equilibrium of the game~$\gameNF{\matx{u}, P_{\tilde{A}_2 | A_2}}$  if
\begin{IEEEeqnarray}{l}
\label{EqBigMaxTontin}
P_{A_2} \in  \arg  \min_{P \in \simplex{\set{A}_2}}\hat{v}\left( P \right) \supersqueezeequ \IEEEeqnarraynumspace \mbox{ and }\\
\label{EqBigMaxTintin}
P_{A_1| \tilde{A}_2}   \in  \BR_1\left( P_{A_2} \right),
\end{IEEEeqnarray}
where the function~$\hat{v}$ is in~\eqref{Eqvhat}, and the correspondence~$ \BR_1$ is in~\eqref{EqBR1}.
\end{definition}
%
%\begin{figure}
%\begin{center}
%\begin{tikzpicture}[scale = 0.6, transform shape,
%squarednodeL/.style={rectangle, draw=black!60, very thick, minimum width=2cm, minimum height=3cm}, 
%squarednode/.style={rectangle, draw=black!60, very thick, minimum width=2cm, minimum height=1.5cm}, 
%squarednode1/.style={rectangle, draw=black!60, very thick,  minimum width=1.5cm, minimum height=1.2cm}, 
%squarednode2/.style={rectangle, draw=black!60, very thick,  minimum width=1.5cm, minimum height=0.6cm}, 
%]
%\node[squarednode]      (P2)                      {\makecell[c]{\bf{Player 2} \\$P_{A_2}$}};
%\node[squarednode]      (noise)          [right=of P2, xshift=0cm]                    {$P_{\tilde{A}_2|A_2}$};
%\node[squarednode]      (P1)          [right=of noise, xshift=0cm]                    {\makecell[c]{\bf{Player 1} \\$P_{A_1|\tilde{A}_2}$}};
%\node[squarednodeL]      (game)          [right=of P1, xshift=0cm, yshift = -0.5cm]                    {$\gameNF{\matx{u}, P_{\tilde{A}_2 | A_2}}$};
%\node[squarednode2, draw=none] (output) [right=of game,xshift=-0.6cm] {$v(P_{A_1|\tilde{A}_2}, P_{A_2})$};
%%
%\draw[->, thick] (P2) -- node [anchor=north, above] {$A_2$}(noise) ;
%\draw[->, thick] (noise) -- node [anchor=north,  above] {$\tilde{A}_2$} (P1) ;
%\draw[->, thick] (P1) -- node [anchor=north,  above] {$A_1$} (game.155) ;
%\draw[->, thick] (P2.east) -- ++(0.5,0) -| ++(0,-0.2) |- (game.215);
%\draw[->, thick] (game) -- (output);
%\end{tikzpicture}
%\end{center}
%\caption{Game $\gameNF{\matx{u}, P_{\tilde{A}_2 | A_2}}$ in \eqref{EqTheNewGame}. }
%\label{FigGameuw}
%\end{figure}
\vspace{-1em}
\section{Preliminaries}\label{SecPrel}
The interest on the game~$\game{G}\left( \matx{u} \right)$ in~\eqref{EqTheGame} stems from the fact that its payoff at the NE is equivalent to the payoff at the equilibrium of the the game~$\game{G}\left(\matx{u}, P_{\tilde{A}_2 | A_2} \right)$ in~\eqref{EqTheNewGame},  under the assumption that  \Pone does not obtain any  information about the action played by \Ptwo from the output of the DMC. That is, $I\left( P_{\tilde{A}_2 | A_2}; P \right) = 0$ for all $P \in \triangle \left( \set{A}_2 \right)$, where~$I\left( \cdot ; \cdot \right)$ is the mutual information.
Let the expected payoff in the game~$\game{G}\left( \matx{u} \right)$ be represented by the function~$u: \simplex{\set{A}_1} \times \simplex{\set{A}_2} \to \reals$ such that, given the strategies~$P_{A_1}$ and~$P_{A_2}$,  
\begin{IEEEeqnarray}{rcl}
\label{EqNormalFormU}
u\left(P_{A_1}, P_{A_2} \right) & = &  \sum_{(i,j) \in \{1,2\}^2} P_{A_1} \left( a_i \right)  P_{A_2}\left(a_j\right) u_{i,j}.
\end{IEEEeqnarray}

The following lemma characterizes the payoff at the NE of the game~$\game{G}\left( \matx{u} \right)$ and shows  that~$2\times2$ ZSGs exhibit either a unique NE or infinitely many NEs. 
\begin{lemma}[Theorem~$1.5$ in \cite{sun:hal-03852615}]\label{LemmaNE}
Let the probability measures $P^{\star}_{A_1} \in \simplex{\set{A}_1}$ and~$P^{\star}_{A_2} \in \simplex{\set{A}_2}$ form a NE of the game $\gameNF{\matx{u}}$ in \eqref{EqTheGame}. 
If the entries of the matrix~$\matx{u}$ in~\eqref{EqMatrixU} satisfy
\begin{subequations}
\begin{IEEEeqnarray}{lcl}
\left( u_{1,1} - u_{1,2} \right) \left( u_{2,2} - u_{2,1} \right)   >   0 & \mbox{ and } & \\
\left( u_{1,1} - u_{2,1} \right) \left( u_{2,2} - u_{1,2} \right)   >   0,
\end{IEEEeqnarray}
\label{EqMixedAssumption}
\end{subequations}
then, the NE of the game~$\gameNF{\matx{u}}$ in~\eqref{EqTheGame} is unique, with 
\begin{subequations}\label{EqNEStratExample}
\begin{IEEEeqnarray}{rcl}
\label{EqPA1StarExample}
P^{\star}_{A_1}(a_1) & = &  \frac{u_{2,2}-u_{2,1}}{u_{1,1} - u_{1,2} - u_{2,1}+u_{2,2}} \in (0,1)\mbox{  and  } \IEEEeqnarraynumspace\\
\label{EqPA2StarExample}
P^{\star}_{A_2}(a_1) & = &\frac{u_{2,2}-u_{1,2}}{u_{1,1} - u_{1,2} - u_{2,1}+u_{2,2}}\in (0,1).
\end{IEEEeqnarray} 
\end{subequations}
Moreover,  the expected payoff at the NE is
\begin{IEEEeqnarray}{rcl}
u(P_{A_1}^{\star},P_{A_2}^{\star}) & = & \frac{u_{1,1}u_{2,2} - u_{1,2}u_{2,1}}{u_{1,1} - u_{1,2} - u_{2,1}+u_{2,2}}.
\end{IEEEeqnarray}
If the entries of the matrix~$\matx{u}$ in~\eqref{EqMatrixU}  satisfy
\begin{subequations}
\begin{IEEEeqnarray}{lcl}
\left( u_{1,1} - u_{1,2} \right) \left( u_{2,2} - u_{2,1} \right)   \leqslant   0 & \mbox{ or } & \\
\left( u_{1,1} - u_{2,1} \right) \left( u_{2,2} - u_{1,2} \right)   \leqslant   0,
\end{IEEEeqnarray}
\label{EqNotMixedAssumption}
\end{subequations}
then, there exists either a unique NE or infinitely many NEs; and all NE strategies lead to the same payoff,
\begin{IEEEeqnarray}{rcl}\label{Equ:f_3}
u(P_{A_1}^{\star},P_{A_2}^{\star}) & = & \min_{j \inCountTwo} \max_{i \inCountTwo} u_{i,j} = \max_{i \inCountTwo} \min_{j \inCountTwo}  u_{i,j}. \IEEEeqnarraynumspace
\end{IEEEeqnarray}
\end{lemma}
A payoff matrix~$\matx{u}$ that satisfies~\eqref{EqMixedAssumption} represents a ZSG exhibiting a unique NE in strictly mixed strategies. Alternatively, a payoff matrix~$\matx{u}$ that satisfies~\eqref{EqNotMixedAssumption} represents a ZSG exhibiting \emph{strategic dominance}, a unique pure NE, or infinitely many NEs~\cite{sun:hal-03852615}. 

Let the function~$\hat{u}: \simplex{\set{A}_2} \to \reals$ be such that for all~$P \in \simplex{\set{A}_2}$, 
\begin{IEEEeqnarray}{rcl}
\label{EqHatu}
\hat{u}\left( P \right) & = & \max_{Q \in \simplex{\set{A}_1}} u\left(Q,  P\right),
\end{IEEEeqnarray}
where the function~$u$ is defined in~\eqref{EqNormalFormU}. 
The function~$\hat{u}$ in~\eqref{EqHatu} determines the payoff~$\hat{u}(P)$ in the game~$\gameNF{\matx{u}}$ in~\eqref{EqTheGame} when \Pone always plays an optimal strategy to the strategy~$P$ played by \Ptwo. 
Moreover, the minimum of the function~$\hat{u}$ is the payoff at the NE.

\section{Main Results} 

\subsection{Characterization of the Equilibria}
The following theorem ensures the existence of an equilibrium for the game $\game{G}\left(\matx{u}, P_{\tilde{A}_2 | A_2} \right)$ in~\eqref{EqTheNewGame}.

\begin{theorem}[Existence]\label{TheoExistance}
The game~$\game{G}\left(\matx{u}, P_{\tilde{A}_2 | A_2} \right)$ in~\eqref{EqTheNewGame} always possesses an equilibrium.
\end{theorem}

\begin{IEEEproof}
The proof is presented in Appendix A of \cite{InriaRR9505}.
\end{IEEEproof}
For characterizing the payoff at the equilibrium of the game $\game{G}\left(\matx{u}, P_{\tilde{A}_2 | A_2} \right)$, it is important to highlight that the set of optimal commitments for \Ptwo are the strategies that minimize the function~$\hat{v}$ in~\eqref{Eqvhat}. 
Let~$P^{(1)}$ and~$P^{(2)}$ be two real numbers such that for all~$i \inCountTwo$,
\begin{IEEEeqnarray}{rcl}
\label{EqPi}
\begin{pmatrix}
1\\
0
\end{pmatrix}^{\sfT}
\matx{u}^{(i)}
\begin{pmatrix}
P^{(i)}\\
1- P^{(i)}
\end{pmatrix} 
& = &
\begin{pmatrix}
0\\
1
\end{pmatrix}^{\sfT}
\matx{u}^{(i)}
\begin{pmatrix}
P^{(i)}\\
1- P^{(i)}
\end{pmatrix},  \IEEEeqnarraynumspace 
\end{IEEEeqnarray}
where the~$2\times2$ matrix~$\matx{u}^{(i)}$ satisfies,
\begin{IEEEeqnarray}{rcl}
\label{EqMatrixUs}
\matx{u}^{(i)} & = & \matx{u}
\begin{pmatrix}
P_{\tilde{A}_2 | A_2 = a_1}(a_i) & 0 \\
 0 & P_{\tilde{A}_2 | A_2 = a_2}(a_i)
\end{pmatrix},
\end{IEEEeqnarray}
with the matrix~$\matx{u}$ defined in~\eqref{EqMatrixU}; and the probability measures~$P_{\tilde{A}_2 | A_2}$ defined in~\eqref{EqTheRabbitChannel}.
Using this notation, the following theorem characterizes the payoff at equilibrium.
\begin{theorem}[Equilibrium Payoff]\label{TheoEquilibrium}
Let the tuple~$\left(P_{A_1| \tilde{A}_2}^{\dagger}, P_{A_2}^{\dagger}\right) \in \simplex{\set{A}_1}^2\times \simplex{\set{A}_2}$ form an equilibrium of the game~$\game{G}\left(\matx{u}, P_{\tilde{A}_2 | A_2} \right)$ in~\eqref{EqTheNewGame}. If the matrix~$\matx{u}$ in~\eqref{EqMatrixU} satisfies~\eqref{EqMixedAssumption}, then
\begin{IEEEeqnarray}{rcl}\label{EqSurgeryPain}
v\left(P_{A_1| \tilde{A}_2}^{\dagger}, P_{A_2}^{\dagger} \right)  & = & \min\lbrace \hat{v}\left( P_{1}\right),\hat{v}\left( P_{2}\right) \rbrace,
\end{IEEEeqnarray}
where, the functions~$v$ and~$\hat{v}$ are defined in~\eqref{EqTheCostFunction} and ~\eqref{Eqvhat}, respectively, and for all~$i \inCountTwo$, the probability measure~$P_i \in \simplex{\set{A}_2}$ is such that~$P_{i}\left( a_1 \right) = P^{(i)}$, with~$P^{(i)}$ in~\eqref{EqPi}.
Alternatively,  if the entries of the matrix~$\matx{u}$ satisfy~\eqref{EqNotMixedAssumption}, then
\begin{IEEEeqnarray}{rcl}
v\left(P_{A_1| \tilde{A}_2}^{\dagger}, P_{A_2}^{\dagger} \right)  & = & \min_{j \inCountTwo} \max_{i \inCountTwo} u_{i,j}. \IEEEeqnarraynumspace
\end{IEEEeqnarray}
\end{theorem}

\begin{IEEEproof}
The proof is presented in Appendix B of \cite{InriaRR9505}.
\end{IEEEproof}
Theorem~\ref{TheoEquilibrium} characterizes the optimal commitment of \Ptwo. More specifically, when the payoff matrix~$\matx{u}$ in~\eqref{EqMatrixU} is such that the game~$\game{G}\left(\matx{u}\right)$ in~\eqref{EqTheGame} possesses a unique NE in mixed strategies (conditions in~\eqref{EqMixedAssumption}), the optimal commitment is one of the strategies~$P_1$ or~$P_{2}$ in~\eqref{EqSurgeryPain}. For all~$i \inCountTwo$, the strategy~$P_{i}$ makes \Pone indifferent to play any of its actions in the game~$\game{G}\left(\matx{u}^{(i)}\right)$, with the matrix~$\matx{u}^{(i)}$ in~\eqref{EqMatrixUs}. This follows from the construction in~\eqref{EqPi}.
Alternatively, when the payoff matrix~$\matx{u}$ in~\eqref{EqMatrixU} is such that the game~$\game{G}\left(\matx{u}\right)$ in~\eqref{EqTheGame} does not possess a unique NE in mixed strategies (conditions in~\eqref{EqNotMixedAssumption}), the optimal commitment for \Ptwo is a pure strategy.  This is equivalent to announcing to \Pone that a given action would be played with probability one, which makes the noisy observation immaterial. 
Moreover, from Lemma~\ref{LemmaNE}, it follows that the payoffs at the NE and the SE in pure strategies of the game~$\game{G}\left(\matx{u}\right)$ are identical to the payoff at the equilibrium of the game~$\game{G}\left(\matx{u}, P_{\tilde{A}_2 | A_2} \right)$. 
That is, neither the fact that \Ptwo commits before its opponent nor the fact that \Pone obtains an observation of the action played by its opponent represent any benefit for either player.

\vspace{-0.5em}
\subsection{The Set of Best Responses of \Pone}
The following lemma shows that, given a commitment~$P_{A_2}$,  the set of best responses~$\BR_1(P_{A_2})$ in~\eqref{EqBR1} is the Cartesian product of two sets that can be independently described. 

\begin{lemma}\label{CorBR1Separates}
The correspondance~$\BR_1$ in~\eqref{EqBR1} satisfies for all~$P \in \simplex{\set{A_2}}$,
\begin{IEEEeqnarray}{rcl}
\label{EqBR1X}
\BR_1\left( P \right) &=& \BR_{1,1}\left( P \right) \times \BR_{1,2}\left( P \right), 
\end{IEEEeqnarray}
where for all~$i \inCountTwo$, the correspondence~$\BR_{1,i}:  \simplex{\set{A}_{2}} \to \mathscr{F}\left(\simplex{\set{A}_1}\right)$ is such that 
\begin{IEEEeqnarray}{rcl}
\label{EqBR1i}
\BR_{1,i}\left( P \right) & = & \arg\max_{ Q \in \simplex{\set{A}_1}} 
\begin{pmatrix}
Q\left( a_1 \right)\\
Q\left( a_2 \right)
\end{pmatrix}^{\sfT}
\matx{u}^{(i)}
\begin{pmatrix}
P\left(a_1\right)\\
P\left(a_2\right)
\end{pmatrix}, \IEEEeqnarraynumspace 
\end{IEEEeqnarray}
where the matrix~$\matx{u}^{(i)}$ is in~\eqref{EqMatrixUs}.
\end{lemma}
\begin{IEEEproof}
The proof is presented in Appendix C of \cite{InriaRR9505}.
\end{IEEEproof}
The following lemma characterizes the sets $\BR_{1,1}\left( P \right)$ and $\BR_{1,2}\left( P \right)$ in~\eqref{EqBR1i}. 
\begin{lemma}\label{LemmaBreakingBad1}
Given a probability measure $P \in \simplex{\set{A}_2}$,  for all $i \inCountTwo$, the correspondence $\BR_{1,i}$ in~\eqref{EqBR1i} satisfies
\begin{IEEEeqnarray}{rCl}\label{Equ:f_1_1}
&&\BR_{1,i}(P)= %\IEEEnonumber \\
%&& 
\ \left\{ 
 \begin{array}{cl}
\hspace{-1.5ex} \{ Q \in \Delta(\set{A}_1): Q(a_1) = 1 \}, &  \hspace{-1ex}\textnormal{if } s_i> 0,\\
\hspace{-1.5ex} \{ Q \in \Delta(\set{A}_1): Q(a_1) = 0 \}, &  \hspace{-1ex}\textnormal{if }  s_i< 0,\\
\simplex{\set{A}_1}, &  \hspace{-1ex}\textnormal{if }  
s_i= 0, 
 \end{array}
 \right. \IEEEeqnarraynumspace \supersqueezeequ
\end{IEEEeqnarray}
where $s_i \in \reals$ is given by 
\begin{IEEEeqnarray}{rcl}
&&s_i \triangleq \left(u_{1,1} - u_{2,1}\right) P\left( a_1 \right) P_{\tilde{A}_2 | A_2 = a_1}\left( a_i \right) \IEEEnonumber \\
&& \quad + \left(u_{1,2}-u_{2,2} \right)P\left( a_2 \right) P_{\tilde{A}_2 | A_2 = a_2}\left( a_i \right). \label{EqBigOne}
\end{IEEEeqnarray}
\end{lemma}
\begin{IEEEproof}
The proof is presented in Appendix D of \cite{InriaRR9505}.
\end{IEEEproof}
A first observation from Lemma~\ref{LemmaBreakingBad1} is that for all~$i \inCountTwo$ and for all~$P \in \simplex{\set{A}_2}$, the cardinality of set~$\BR_{1,i}\left( P \right)$ is either one or infinite. In the case in which~$\BR_{1,i}\left( P \right)$ is a singleton, the only element is a pure strategy. Alternatively, when the cardinality is infinity, the set~$\BR_{1,i}\left( P \right)$ is identical to the set of all possible probability measures on~$\set{A}_1$, i.e.,~$\BR_{1,i}\left( P \right) = \simplex{\set{A}_1}$. That is, \Pone chooses its actions either indifferently (all strategies are best responses) or deterministically (pure strategy). This contrasts with the case of bi-matrix Stakelberg games in which the existence of multiple best responses constraints the existence of equilibria \cite{lucchetti_1987_existence}. 

Note also that  for all~$\left( i, j\right) \inCountTwo^2$, the expected payoff, when \Pone plays~$a_j$, \Ptwo has committed to~$P_{A_2}$, and the noisy observation is~$a_i$, is~$u_{j,1} P_{A_2}\left( a_1 \right) P_{\tilde{A}_2 | A_2 = a_1}\left( a_i \right) \supersqueezeequ$ $+$ $u_{j,2} P_{A_2}\left( a_2 \right) P_{\tilde{A}_2 | A_2 = a_2}\left( a_i \right)\supersqueezeequ$. 
Thus, the right-hand side of the equality in~\eqref{EqBigOne} is the difference between the expected payoff obtained when \Pone plays~$a_1$ and when it plays~$a_2$, subject to the observation~$a_i$ and the commitment~$P_{A_2}$. 

The following lemma presents a different view of the correspondences~$\BR_{1,1}$ and~$\BR_{1,2}$ in~\eqref{EqBR1i}. It suggests that  \Pone performs an estimation of the likelihood with which \Ptwo might have played each of its actions based on the knowledge of the commitment and the noisy observation.

\begin{lemma}\label{LemmaBreakingGrounds}
Given a probability measure $P \in \simplex{\set{A}_2}$,  for all $i \inCountTwo$, the correspondence $\BR_{1,i}$ in~\eqref{EqBR1i} satisfies
\begin{IEEEeqnarray}{rcl}
\BR_{1,i}\left( P \right) & = & \arg
\max_{Q \in \simplex{\set{A}_1}} u\left( Q , P_{A_2 | \tilde{A}_2 = a_i} \right),
\end{IEEEeqnarray}
where the function $u$ is defined in~\eqref{EqNormalFormU}; the probability measure $P_{A_2 | \tilde{A}_2 = a_i}$ satisfies for all $j \inCountTwo$,
\begin{IEEEeqnarray}{rcl}
\label{EqPost}
P_{A_2 | \tilde{A}_2 = a_i} \left( a_j \right) & = & \frac{P_{\tilde{A}_2 | A_2 = a_j} \left( a_i \right) P\left( a_j \right)}{\displaystyle\sum_{\ell \inCountTwo} P_{\tilde{A}_2 | A_2 = a_{\ell}} \left( a_i \right) P\left( a_{\ell} \right)},
\end{IEEEeqnarray}
with the probability measures $P_{\tilde{A}_2 | A_2 = a_{1}}$ and $P_{\tilde{A}_2 | A_2 = a_{2}}$ defined in~\eqref{EqTheRabbitChannel}.
\end{lemma}
\begin{IEEEproof}
The proof is presented in Appendix E of \cite{InriaRR9505}.
\end{IEEEproof}

For all~$(i,j) \inCountTwo^2$,  the likelihood with which \Ptwo has chosen action~$a_j$ given the commitment~$P$ and the noisy observation~$a_i$ is~$P_{A_2 | \tilde{A}_2 = a_i} \left( a_j \right)$ in~\eqref{EqPost}. 
Hence, from Lemma~\ref{LemmaBreakingBad1} and Lemma~\ref{LemmaBreakingGrounds}, the optimal strategy of \Pone to the observation~$a_i$ and the commitment~$P$ in the game~$\gameNF{\matx{u}, P_{\tilde{A}_2 | A_2}}$ is identical to its optimal strategy in the game~$\gameNF{\matx{u}}$ in~\eqref{EqTheGame} when its  opponent plays the strategy~$P_{A_2 | \tilde{A}_2 = a_i}$  in~\eqref{EqPost}. 

\subsection{Relevance of Noisy Observations}

The following lemma shows that the function~$\hat{u}$ in~\eqref{EqHatu} is upper bounded by the function~$\hat{v}$ in~\eqref{Eqvhat}. This implies that, granting observations to \Pone of the actions played by \Ptwo does not harm \Pone. On the contrary, in some cases it might significantly benefit it.  
%
%\begin{lemma}\label{LemmaLowerBounds}
%Let the probability measures $P^{\star}_{A_1} \in \simplex{\set{A}_1}$ and $P^{\star}_{A_2} \in \simplex{\set{A}_2}$ form one of the NEs of the game $\gameNF{\matx{u}}$ in~\eqref{EqTheGame}. 
%Let also the tuple $\left(P_{A_1| \tilde{A}_2}^{\dagger}, P_{A_2}^{\dagger}\right) \in \simplex{\set{A}_1}^2\times \simplex{\set{A}_2}$ form an equilibrium of the game $\game{G}\left(\matx{u}, P_{\tilde{A}_2 | A_2} \right)$ in~\eqref{EqTheNewGame}. 
%For all $P \in \simplex{\set{A}_2}$, it holds that
%\begin{IEEEeqnarray}{rcl}\label{EqLBTrickySurgery}
%%&&
% u(P_{A_1}^{\star},P_{A_2}^{\star}) 
% & \leqslant & v\left(P_{A_1| \tilde{A}_2}^{\dagger}, P_{A_2}^{\dagger} \right) \\
%\label{EqLBTrickyB}
% & \leqslant &  \sum_{k\inCountTwo}P(a_k) \left( \max_{i\inCountTwo}  u_{i,k} \right),  \IEEEeqnarraynumspace \supersqueezeequ
%\end{IEEEeqnarray}
%where the functions $\hat{v}$, $u$,  and $\hat{u}$ are defined in~\eqref{Eqvhat},~\eqref{EqNormalFormU},   and~\eqref{EqHatu}, respectively.
%\end{lemma}
\begin{lemma}\label{LemmaLowerBounds}
Let the probability measures~$P^{\star}_{A_1} \in \simplex{\set{A}_1}$ and~$P^{\star}_{A_2} \in \simplex{\set{A}_2}$ form one of the NEs of the game~$\gameNF{\matx{u}}$ in~\eqref{EqTheGame}. 
For all~$P \in \simplex{\set{A}_2}$, it holds that
\begin{IEEEeqnarray}{c}\label{EqLBTrickySurgery}
%&&
 u(P_{A_1}^{\star},P_{A_2}^{\star}) 
  \leqslant  \hat{u}(P)   \leqslant  \hat{v}(P) 
  \leqslant   \sum_{j\inCountTwo}P(a_j) \left( \max_{i\inCountTwo}  u_{i,j} \right),  \IEEEeqnarraynumspace \Tsupersqueezeequ
\end{IEEEeqnarray}
where the functions~$\hat{v}$,~$u$,  and~$\hat{u}$ are defined in~\eqref{Eqvhat},~\eqref{EqNormalFormU},   and~\eqref{EqHatu}, respectively.
\end{lemma}
\begin{IEEEproof}
The proof is presented in Appendix F of \cite{InriaRR9505}.
\end{IEEEproof}

The following lemma compares the payoffs at the equilibria of the games~$\gameNF{\matx{u}}$ in~\eqref{EqTheGame} and~$\game{G}(\matx{u}, P_{\tilde{A}_2 | A_2})$  in~\eqref{EqTheNewGame}.

\begin{lemma}\label{LemmaCompareEquilibria}
Let the probability measures~$P^{\star}_{A_1} \in \simplex{\set{A}_1}$ and~$P^{\star}_{A_2} \in \simplex{\set{A}_2}$ form one of the NEs of the game~$\gameNF{\matx{u}}$ in~\eqref{EqTheGame}. 
Let also the tuple~$\left(P_{A_1| \tilde{A}_2}^{\dagger}, P_{A_2}^{\dagger}\right) \in \simplex{\set{A}_1}^2\times \simplex{\set{A}_2}$ form an equilibrium of the game~$\game{G}\left(\matx{u}, P_{\tilde{A}_2 | A_2} \right)$ in~\eqref{EqTheNewGame}. 
 Then,
\begin{IEEEeqnarray}{rcl}
u(P_{A_1}^{\star},P_{A_2}^{\star})  \leqslant v\left(P^{\dagger}_{A_1|\tilde{A}_2}, P^{\dagger}_{A_2}\right) \leqslant \min_{j\in\{1,2\}}\max_{i\in\{1,2\}}  u_{i,j}. \IEEEeqnarraynumspace
\end{IEEEeqnarray}
\end{lemma}
\begin{IEEEproof}
The proof is presented in Appendix G of \cite{InriaRR9505}.
\end{IEEEproof}

Lemma~\ref{LemmaCompareEquilibria} reveals that the payoff at the equilibria of the game $\game{G}(\matx{u}, P_{\tilde{A}_2 | A_2})$  in~\eqref{EqTheNewGame} is lower bounded by the NE of the game $\gameNF{\matx{u}}$ in~\eqref{EqTheGame}, which coincides with the SE in mixed strategies; and is upper bounded by the SE in pure strategies of the game $\gameNF{\matx{u}}$. The lower bound corresponds to the case in which the \Pone does not observe the actions of its opponent, while the upper bound corresponds to the case in which \Pone has perfect observations of the actions taken by \Ptwo.

The following lemma presents necessary and sufficient conditions under which the payoff at the  equilibrium of the game~$\game{G}\left(\matx{u}, P_{\tilde{A}_2 | A_2}  \right)$ is not greater than the NE of the game~$\game{G}\left(\matx{u} \right)$.

\vspace{-0.5em}
\begin{lemma}\label{LemmaEquilibriumEquality1}
Let the tuple~$\left( P_{A_1| \tilde{A}_2}^{\dagger}, P_{A_2}^{\dagger}\right) \in \simplex{\set{A}_1}^2 \times \simplex{\set{A}_2}$ form an equilibrium of the game~$\game{G}\left(\matx{u}, P_{\tilde{A}_2 | A_2} \right)$ in~\eqref{EqTheNewGame}. Let also the tuple~$\left(P^{\star}_{A_1}, P^{\star}_{A_2}\right) \in \simplex{\set{A}_1} \times \simplex{\set{A}_2}$ form one of the NEs of the game~$\gameNF{\matx{u}}$ in~\eqref{EqTheGame}. Then, 
\begin{IEEEeqnarray}{rcl}\label{Equ:f_5}
v\left( P_{A_1| \tilde{A}_2}^ \dagger, P_{A_2}^ \dagger\right)  & = &u(P_{A_1}^{\star},P_{A_2}^{\star})  , 
\end{IEEEeqnarray}
if and only if, 
$(a)$ the matrix~$\matx{u}$ in~\eqref{EqMatrixU} satisfies~\eqref{EqNotMixedAssumption}; or 
$(b)$ the matrix~$\matx{u}$ in~\eqref{EqMatrixU} satisfies~\eqref{EqMixedAssumption} and the DMC in~\eqref{EqTheRabbitChannel} satisfies~for all $P \in \triangle \left( \set{A}_2 \right)$, that $I\left( P_{\tilde{A}_2 | A_2}; P \right)~=~0$.
\end{lemma}
\begin{IEEEproof}
The proof is presented in Appendix H of \cite{InriaRR9505}.
\end{IEEEproof}
Lemma~\ref{LemmaEquilibriumEquality1} establishes that  granting \Pone with noisy observations of the action played by \Ptwo does not make any difference in two particular scenarios. 
First, in ZSGs with strategic dominance, NEs in pure strategies and infinitely many NEs (condition~$(a)$).
Second, in ZSGs when the DMC in~\eqref{EqTheRabbitChannel} is such that \Pone does not obtain any information about the action played by \Ptwo by observing the output of the DMC. 

Lemma~\ref{LemmaLowerBounds} and Lemma~\ref{LemmaEquilibriumEquality1} imply that granting \Pone with relevant noisy observations of the action played by \Ptwo makes a difference exclusively for ZSGs with a unique NE in mixed strategies. In this case, given  the commitment of the leader $P_{A_2}$,  relevant noisy observations refer to observations obtained through a DMC exhibiting positive mutual information between the channel input and the channel output. That is,~$I\left( P_{\tilde{A}_2 | A_2}; P_{A_2} \right) > 0$. 

The following lemma describes a special class of channels.

\begin{lemma}\label{LemmaPureEquilibriumEquality}
Let~$\left(P_{A_1| \tilde{A}_2}^{\dagger}, P_{A_2}^{\dagger}\right) \in \simplex{\set{A}_1}^2 \times \simplex{\set{A}_2}$ form an equilibrium of the game~$\game{G}\left(\matx{u}, P_{\tilde{A}_2 | A_2} \right)$ in~\eqref{EqTheNewGame}.  If~ for all $P \in \triangle \left( \set{A}_2 \right)$,~$I\left( P_{\tilde{A}_2 | A_2}; P \right) = H\left( P \right) = H\left( P_{\tilde{A}_2} \right)$, with $P_{\tilde{A}_2}(a_i) = \sum_{\ell \inCountTwo} P_{\tilde{A}_2 | A_2 = a_{\ell}} \left( a_i \right) P\left( a_{\ell} \right)$ and  $i \inCountTwo$, then
\begin{IEEEeqnarray}{rcl}\label{Equ:PureEquilibriumEquality}
\hat{v}\left( P_{A_2}^ \dagger\right) & = & \min_{j \inCountTwo} \max_{i \inCountTwo} u_{i,j}. \IEEEeqnarraynumspace
\end{IEEEeqnarray}
\end{lemma}

\begin{IEEEproof}
The proof is presented in Appendix I of \cite{InriaRR9505}.
\end{IEEEproof}
 The condition that for all $P \in \triangle \left( \set{A}_2 \right)$,~$I\left( P_{\tilde{A}_2 | A_2}; P \right) = H\left( P \right) = H\left( P_{\tilde{A}_2} \right)$ implies that the DMC in~\eqref{EqTheRabbitChannel} establishes a deterministic bijection between the channel input and the channel output. 
From this perspective, Lemma~\ref{LemmaPureEquilibriumEquality}  strengthens the observation that under perfect observations of the action played by \Ptwo, the commitment becomes irrelevant and the payoff at the equilibrium of the game~$\game{G}\left(\matx{u}, P_{\tilde{A}_2 | A_2} \right)$ in~\eqref{EqTheNewGame} is identical to the SE in pure strategies of the game~$\game{G}\left(\matx{u}\right)$ in~\eqref{EqTheGame}, i.e., the~$\min\max$ solution in pure strategies.  

\begin{figure}[t!]
\centering
        \begin{subfigure}[b]{0.475\textwidth}
            \centering
            \includegraphics[width=\textwidth]{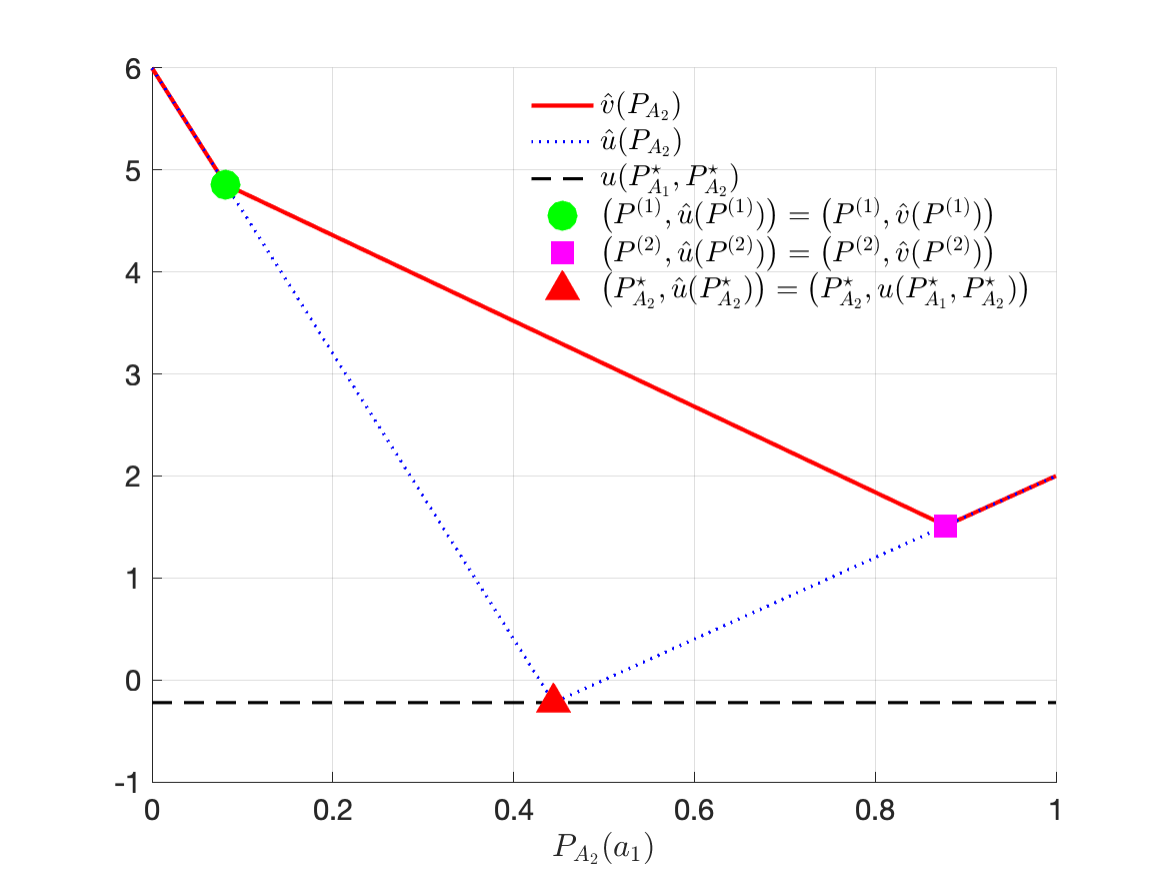}
            \caption{Payoff matrix $\matx{u} =\left( -8, 6 ; 2, -2\right)$ in \eqref{EqMatrixU}.}    
        \end{subfigure}
        \hfill
        \begin{subfigure}[b]{0.475\textwidth}  
            \centering 
            \includegraphics[width=\textwidth]{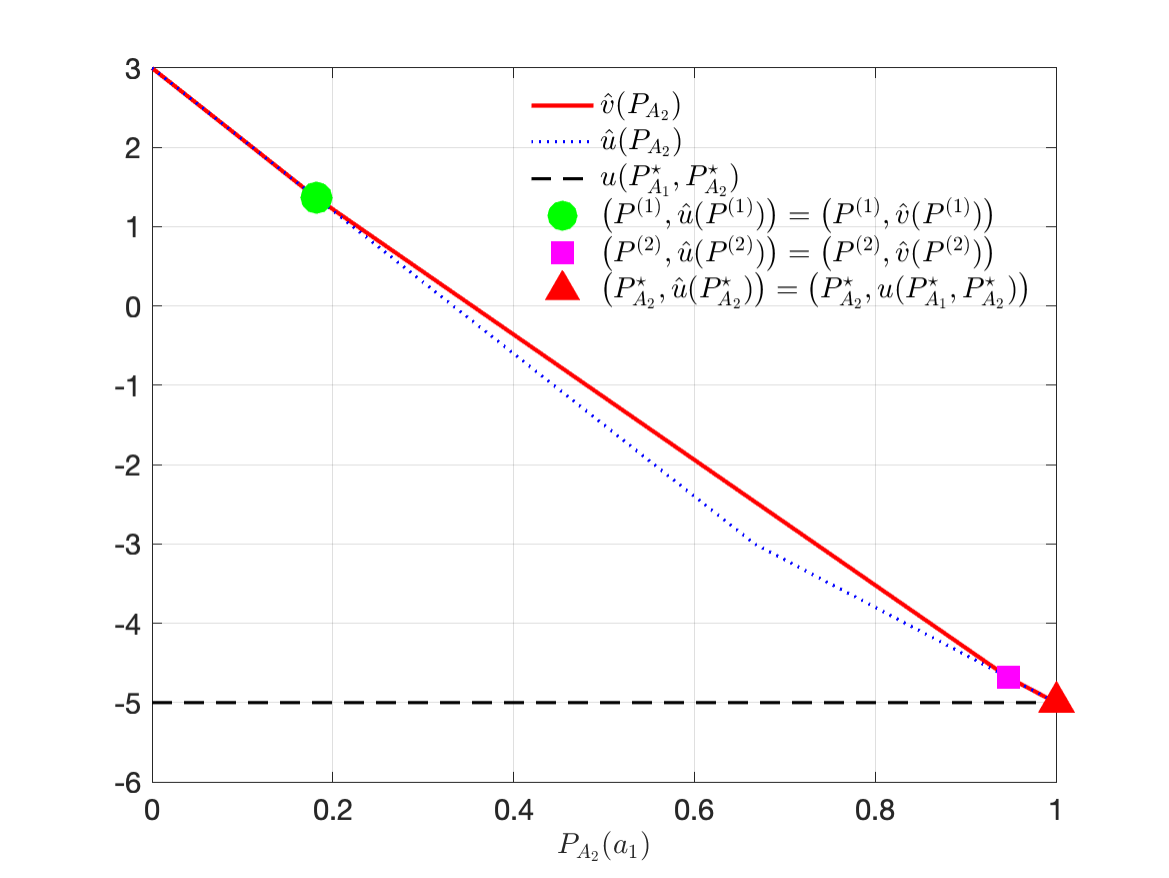}
            \caption{Payoff matrix $\matx{u} = \left( -5, 1 ; -6, 3\right)$ in \eqref{EqMatrixU}.}    
        \end{subfigure}
%        \vskip\baselineskip
%        \begin{subfigure}[b]{0.475\textwidth}   
%            \centering 
%            \includegraphics[width=\textwidth]{Figures/3.pdf}
%            \caption{Payoff matrix $\matx{u} =\left( 2, 9 ; -9, 5\right)$ and  channel matrix $\matx{w} = \left( 0.9,  0.1; 0.1, 0.9\right)$, with $P^{(1)} = $, $P^{(2)} = $, and $P_{A_2}^{\star}\left( a_1 \right) = $.}    
%        \end{subfigure}
%        \hfill
%        \begin{subfigure}[b]{0.475\textwidth}   
%            \centering 
%            \includegraphics[width=\textwidth]{Figures/4.pdf}
%            \caption{Payoff matrix $\matx{u} =\left( 5, 5 ; -6, 10\right)$ and  channel matrix $\matx{w} = \left( 0.9,  0.1; 0.1, 0.9\right)$, with $P^{(1)} = $, $P^{(2)} = $, and $P_{A_2}^{\star}\left( a_1 \right) \in \left[X, 1 \right] $.}    
%        \end{subfigure}
\caption{Plots of the function $\hat{v}$ in \eqref{Eqvhat} and $\hat{u}$ in \eqref{EqHatu} as a function of the commitment $P_{A_2}$ of \Ptwo with  a symmetric DMC $P_{\tilde{A}_2 | A_2 = a_{1}}(a_1) = P_{\tilde{A}_2 | A_2 = a_{2}}(a_2) =  0.9$ in \eqref{EqTheRabbitChannel}.}
\label{FigBellesFigures}
\end{figure} 

%Figure \ref{FigBellesFigures} depicts the functions $\hat{v}$ in \eqref{Eqvhat} and $\hat{u}$ in \eqref{EqHatu} as a function of the strategy $P_{A_2}$ of \Ptwo. The payoffs in the games  $\game{G}\left(\matx{u}, P_{\tilde{A}_2 | A_2} \right)$ in~\eqref{EqTheNewGame} and  $\game{G}\left(\matx{u}\right)$ in~\eqref{EqTheGame} are respectively $\hat{v}\left(P_{A_2} \right)$ and $\hat{u}\left(P_{A_2} \right)$. The minima of such functions are respectively the payoffs at the corresponding equilibria in each game.
\section{Examples}
In Figure~\ref{FigBellesFigures}(a), the matrix~$\matx{u} =\left( -8, 6 ; 2, -2\right)$ is such that the game~$\game{G}\left(\matx{u}\right)$ exhibits a unique NE in mixed strategies (Lemma~\ref{LemmaNE}). Hence, as announced by Lemma~\ref{LemmaLowerBounds} and Lemma~\ref{LemmaEquilibriumEquality1}, there exists a strict inequality between the NE payoff~$u(P_{A_1}^{\star},P_{A_2}^{\star})$  of the game~$\game{G}\left(\matx{u}\right)$ (red triangle) and the equilibrium payoff~$v\left(P_{A_1| \tilde{A}_2}^{\dagger}, P_{A_2}^{\dagger} \right)$ of the game~$\game{G}\left(\matx{u}, P_{\tilde{A}_2 | A_2} \right)$ (magenta square).
Alternatively, in Figure~\ref{FigBellesFigures}(b), the matrix~$\matx{u} = \left( -5, 1 ; -6, 3\right)$ is such that the game~$\game{G}\left(\matx{u}\right)$ exhibits a unique NE in pure strategies (Lemma~\ref{LemmaNE}). Hence, as predicted by Lemma~\ref{LemmaEquilibriumEquality1}, the payoffs of the games $\game{G}\left(\matx{u}\right)$ and $\game{G}\left(\matx{u}, P_{\tilde{A}_2 | A_2} \right)$ are identical (red triangle). That is,~$u(P_{A_1}^{\star},P_{A_2}^{\star})$  $=$ $v\left(P_{A_1| \tilde{A}_2}^{\dagger}, P_{A_2}^{\dagger} \right)$.

\IEEEtriggeratref{13}
\bibliographystyle{IEEEtran}
\bibliography{reference_SunPerlaza}

\begin{thebibliography}{10}
\providecommand{\url}[1]{#1}
\csname url@samestyle\endcsname
\providecommand{\newblock}{\relax}
\providecommand{\bibinfo}[2]{#2}
\providecommand{\BIBentrySTDinterwordspacing}{\spaceskip=0pt\relax}
\providecommand{\BIBentryALTinterwordstretchfactor}{4}
\providecommand{\BIBentryALTinterwordspacing}{\spaceskip=\fontdimen2\font plus
\BIBentryALTinterwordstretchfactor\fontdimen3\font minus
  \fontdimen4\font\relax}
\providecommand{\BIBforeignlanguage}[2]{{%
\expandafter\ifx\csname l@#1\endcsname\relax
\typeout{** WARNING: IEEEtran.bst: No hyphenation pattern has been}%
\typeout{** loaded for the language `#1'. Using the pattern for}%
\typeout{** the default language instead.}%
\else
\language=\csname l@#1\endcsname
\fi
#2}}
\providecommand{\BIBdecl}{\relax}
\BIBdecl

\bibitem{nash_1950_equilibrium}
J.~F. Nash, ``Equilibrium points in n-person games,'' \emph{Proceedings of the
  National Academy of Sciences}, vol.~36, no.~1, pp. 48--49, 1950.

\bibitem{Stackelberg-1952}
H.~Stackelberg, \emph{Theory of the Market Economy}.\hskip 1em plus 0.5em minus
  0.4em\relax Oxford University Press, Mar. 1952.

\bibitem{conitzer_2006_computing}
V.~Conitzer and T.~Sandholm, ``Computing the optimal strategy to commit to,''
  in \emph{Proc. ACM Conf. on Electronic Commerce}, Ann Arbor, Michigan, USA,
  Jun. 2006, pp. 82--90.

\bibitem{conitzer_2016_stackelberg}
V.~Conitzer, ``On {S}tackelberg mixed strategies,'' \emph{Synthese}, vol. 193,
  no.~3, pp. 689--703, Mar. 2016.

\bibitem{leonardos_2018_commitment}
S.~Leonardos and C.~Melolidakis, ``On the commitment value and commitment
  optimal strategies in bimatrix games,'' \emph{International Game Theory
  Review}, vol.~20, no.~3, p. 1840001, Sep. 2018.

\bibitem{von_2010_leadership}
B.~von Stengel and S.~Zamir, ``Leadership games with convex strategy sets,''
  \emph{Games and Economic Behavior}, vol.~69, no.~2, pp. 446--457, Jul. 2010.

\bibitem{v1928theorie}
J.~v.~Neumann, ``Zur {T}heorie der {G}esellschaftsspiele,'' \emph{Mathematische
  annalen}, vol. 100, no.~1, pp. 295--320, 1928.

\bibitem{simaan_1973_stackelberg}
M.~Simaan and J.~B. Cruz, ``On the {S}tackelberg strategy in nonzero-sum
  games,'' \emph{Journal of Optimization Theory and Applications}, vol.~11,
  no.~5, pp. 533--555, May 1973.

\bibitem{simaan_1973_additional}
------, ``Additional aspects of the {S}tackelberg strategy in nonzero-sum
  games,'' \emph{Journal of Optimization Theory and Applications}, vol.~11,
  no.~6, pp. 613--626, 1973.

\bibitem{jin_2020_local}
C.~Jin, P.~Netrapalli, and M.~Jordan, ``What is local optimality in
  nonconvex-nonconcave minimax optimization?'' in \emph{Proc. Int. Conf. on
  Machine Learning}, Vritual, Jul. 2020, pp. 4880--4889.

\bibitem{bai_2021_sample}
Y.~Bai, C.~Jin, H.~Wang, and C.~Xiong, ``Sample-efficient learning of
  {S}tackelberg equilibria in general-sum games,'' in \emph{Proc. Advances in
  Neural Information Processing Systems}, vol.~34, Virtual, Dec. 2021, pp.
  25\,799--25\,811.

\bibitem{wallmeier_1988_games}
H.~W. Wallmeier, ``Games with informants: {A}n information-theoretical approach
  towards a game-theoretical problem,'' \emph{International Journal of Game
  Theory}, vol.~17, no.~4, pp. 245--278, 1988.

\bibitem{bagwell_1995_commitment}
K.~Bagwell, ``Commitment and observability in games,'' \emph{Games and Economic
  Behavior}, vol.~8, no.~2, pp. 271--280, 1995.

\bibitem{muller_2001_quality}
W.~M{\"u}ller, ``The quality of the signal matters: {A} note on imperfect
  observability and the timing of moves,'' \emph{Journal of Economic Behavior
  \& Organization}, vol.~45, no.~1, pp. 99--106, Mar. 2001.

\bibitem{ferreira_2011_note}
J.~Ferreira, ``A note on {B}agwell's paradox and forward induction in three
  classic games,'' \emph{International Game Theory Review}, vol.~13, no.~01,
  pp. 93--104, 2011.

\bibitem{teng_2013_generalized}
J.~Teng, ``A generalized {S}tackelberg model with noisy observability and
  incomplete information,'' \emph{Available at SSRN 2265130}, 2013.

\bibitem{van_1997_games}
E.~van Damme and S.~Hurkens, ``Games with imperfectly observable commitment,''
  \emph{Games and Economic Behavior}, vol.~21, no. 1-2, pp. 282--308, 1997.

\bibitem{guth_2006_noisy}
W.~G{\"u}th, W.~M{\"u}ller, and Y.~Spiegel, ``Noisy leadership: An experimental
  approach,'' \emph{Games and Economic Behavior}, vol.~57, no.~1, pp. 37--62,
  Oct. 2006.

\bibitem{bizzotto_2022_limits}
J.~Bizzotto, T.~Hinnosaar, and A.~Vigier, ``The limits of commitment,''
  \emph{arXiv preprint arXiv:2205.05546}, 2022.

\bibitem{adolph_1996_commitment}
B.~Adolph, ``Commitment, trembling hand imperfection and observability in
  games,'' \emph{Available at SSRN 2159}, 1996.

\bibitem{Shannon-1948a}
C.~E. Shannon, ``A mathematical theory of communication,'' \emph{The Bell
  System Technical Journal}, vol.~27, pp. 379--423, Jul. 1948.

\bibitem{Shannon-1948b}
------, ``A mathematical theory of communication,'' \emph{The Bell System
  Technical Journal}, vol.~27, pp. 623--656, Oct. 1948.

\bibitem{sun:hal-03852615}
K.~Sun, ``{Some Properties of the Nash Equilibrium in 2 x 2 Zero-Sum Games},''
  {INRIA}, Centre Inria d’Université Côte d’Azur, Sophia Antipolis,
  France, Tech. Rep. RR-9492, Nov. 2022.

\bibitem{InriaRR9505}
K.~Sun, S.~M. Perlaza, and A.~Jean-Marie, ``$2~\times~2$ zero-sum games with
  commitments and noisy observations,'' {INRIA}, Centre Inria d’Université
  Côte d’Azur, Sophia Antipolis, France, Tech. Rep. RR-9505, May 2023.

\bibitem{lucchetti_1987_existence}
R.~Lucchetti, F.~Mignanego, and G.~Pieri, ``Existence theorems of equilibrium
  points in {S}tackelberg,'' \emph{Optimization}, vol.~18, no.~6, pp. 857--866,
  1987.

\end{thebibliography}
%%%
 
 \end{document}